%% file: main.tex
\documentclass[a4paper,11pt]{article}
\pdfoutput=1
\usepackage[utf8]{inputenc}
\usepackage{jheppub}
\usepackage[T1]{fontenc}
\usepackage{amsmath}

\usepackage{graphicx}
\usepackage{color}
\usepackage{amsmath}
\usepackage{graphicx,textcomp,float,gensymb,wrapfig, enumitem,comment,dsfont,subfigure,framed,slashed,appendix,wrapfig,wasysym}
\usepackage{comment}
\usepackage[export]{adjustbox}
\usepackage[utf8]{inputenc}

\newcommand{\gs}{g_\star}
\newcommand{\gss}{g_{\star \mathfrak{s}}}
\newcommand{\Trh}{T_\text{rh}}

\newcommand{\be}{\begin{equation}}
\newcommand{\ee}{\end{equation}}
\newcommand{\bea}{\begin{eqnarray}}  
\newcommand{\eea}{\end{eqnarray}}

\title{FIMP Dark Matter\\in Clockwork/Linear Dilaton Extra-Dimensions}

\author[a]{Nicolás Bernal,$^1$\note{ORCID: \href{https://orcid.org/0000-0003-1069-490X}{0000-0003-1069-490X}}}
\author[b]{Andrea Donini,$^2$\note{ORCID: \href{https://orcid.org/0000-0001-6668-5477}{0000-0001-6668-5477}}}
\author[b]{Miguel G. Folgado,$^3$\note{ORCID: \href{https://orcid.org/0000-0003-1613-5500}{0000-0003-1613-5500}}}
\author[b]{Nuria Rius.$^4$\note{ORCID: \href{http://orcid.org/0000-0002-0606-4297}{0000-0002-0606-4297}}}

\affiliation[a]{Centro de Investigaciones, Universidad Antonio Nariño,\\
Carrera 3 Este \# 47A-15, Bogotá, Colombia}
\affiliation[b]{Instituto de Física Corpuscular, Universidad de Valencia and CSIC,\\
Edificio Institutos Investigación, Catedrático Jose Beltrán 2, Paterna, 46980 Spain}

\emailAdd{nicolas.bernal@uan.edu.co}
\emailAdd{donini@ific.uv.es}
\emailAdd{\\migarfol@ific.uv.es}
\emailAdd{nuria.rius@ific.uv.es}

\abstract{We study the possibility that Dark Matter (DM) is made of Feebly Interacting Massive Particles (FIMP) interacting just gravitationally with the Standard Model particles in the 
framework of a Clockwork/Linear Dilaton (CW/LD) model. We restrict here to the case in which the DM particles are scalar fields. 
This paper extends our previous study of FIMP's in Randall-Sundrum (RS) warped extra-dimensions. As it was the case in the RS scenario,  also in the CW/LD model we find a significant region of the parameter space in which the observed DM relic abundance can be reproduced with scalar DM mass in the MeV range, with a reheating temperature varying from 10 GeV to  $10^{9}$ GeV. We comment on  the similarities of the results in both extra-dimensional models.
}

\begin{document}

\begin{flushright}
PI/UAN-2020-684FT\\
FTUV-20-0908.8746\\
IFIC/20-43
\end{flushright}

\maketitle

\section{Introduction}

Despite the undeniable success of the  Standard Model (SM)  of particle physics, we have   multiple hints pointing to the fact that it cannot be the complete theory of Nature:
from the experimental side there are several observations that the SM does not explain (for instance, the dark matter and dark energy of the Universe, the generation of the matter-antimatter asymmetry or the origin of neutrino masses), 
and also from the theoretical side the SM seems incomplete, as it contains a large number of arbitrary parameters,  lacks of an explanation for the pattern of lepton and quark masses, and suffers from a hierarchy problem. In this work we address the dark matter puzzle in the context of a particular solution to the hierarchy problem, namely the ClockWork/Linear Dilaton (CW/LD) model~\cite{Antoniadis:2011qw, Cox:2012ee,Giudice:2016yja, Giudice:2017fmj}.

Unveiling the nature of dark matter (DM)  is one of the most intriguing challenges in fundamental physics at present. 
Given that all current manifestations of  its existence are of gravitational origin, it is conceivable that these are the only interactions of the DM particles with the SM ones, in which case all experimental efforts for detecting them by other means will not succeed. There is however a way out: a well known theoretical  solution to the hierarchy or naturalness problem
 ({\em i.e.}, the huge hierarchy between the electroweak scale, $\Lambda_{\rm EW} \sim 200$ GeV, and the Planck mass, $M_P \sim 10^{19}$ GeV)
 is to assume that there are extra spatial dimensions. Depending on the particular extra-dimensional scenario, the problem is solved because the fundamental gravitational scale in $D$ dimensions  is $M_D \ll M_P $ 
 (as in the case of Large Extra Dimensions (LED)~\cite{Antoniadis:1990ew, Antoniadis:1997zg, ArkaniHamed:1998rs,Antoniadis:1998ig,ArkaniHamed:1998nn}),
or due to a warping of the space-time, which induces an effective Planck scale  in the four-dimensional brane $\Lambda \ll M_P$ 
(as in  Randall-Sundrum models (RS)~\cite{Randall:1999ee,Randall:1999vf}), or by a mixture of the two mechanisms (as it occurs in the CW/LD model~\cite{Antoniadis:2011qw, Cox:2012ee,Giudice:2016yja, Giudice:2017fmj}). 
All of these models rephrase the hierarchy problem (namely, the existence of two very different mass scales in the SM) into a geometrical problem (for example, in the context of LED, there is no dynamical mechanism that 
may explain why the size of the extra-dimensions be so large, or $R_{\rm extra} \gg 1/\Lambda_{\rm EW}$). A common feature of extra-dimensional models is, however, that
in any case the gravitational interaction is enhanced and thus a DM particle with just such interaction could become detectable.  

Gravity-mediated DM WIMP's ({\em i.e.}, stable or cosmologically  long-lived weakly interactive massive particle, with mass typically in  the range 100 - 1000 GeV, and whose relic abundance is set via the freeze-out mechanism) have been studied mainly in the framework of the RS scenario~\cite{Lee:2013bua, Lee:2014caa, Han:2015cty, Rueter:2017nbk, Rizzo:2018ntg, Carrillo-Monteverde:2018phy, Rizzo:2018joy, Brax:2019koq, Folgado:2019sgz, Goyal:2019vsw}, 
with generic spin-2 mediators~\cite{Kang:2020huh,Chivukula:2020hvi,Kang:2020yul,Kang:2020afi}
and in the context of the CW/LD model~\cite{Folgado:2019gie}.

However, the strength of the DM gravitational interaction could be tiny enough that DM particles are feebly interacting massive particles (FIMP's) and,  thus, 
the relic abundance of DM is set via the  DM freeze-in production mechanism~\cite{McDonald:2001vt, Choi:2005vq, Kusenko:2006rh, Petraki:2007gq, Hall:2009bx, Barman:2020plp}, instead  
(for a recent review see Ref.~\cite{Bernal:2017kxu}).
This alternative possibility has only been studied in the RS model~\cite{Brax:2019koq, Bernal:2020fvw}, so here we extend the analysis of gravity-mediated FIMP DM in extra dimensions to the CW/LD scenario.  As in our previous work~\cite{Bernal:2020fvw},  we consider  ultraviolet (UV) freeze-in~\cite{Elahi:2014fsa} for which the temperature of the thermal bath is always lower than the scale of new physics, 
which in our model is the  fundamental Planck scale, $M_5$, at which the gravitational sector becomes strongly interacting. We also adopt a  phenomenological approach, 
and allow $M_5$ to vary in a wide range ($M_5 \in [10^2, 10^{16}]$~GeV) to fully explore 
the parameter space that could lead to the correct DM relic abundance via freeze-in, highlighting the regions in which it  is also possible to solve the hierarchy problem.

To our knowledge, this is the first study of  freeze-in DM in the continuous (extra-dimensional) clockwork scenario. 
Notice, though,  that infrared (IR) DM freeze-in through the Higgs portal has been analyzed in several discrete clockwork models~\cite{Kim:2017mtc, Kim:2018xsp,Goudelis:2018xqi}.

The paper is organized as follows: in Sec.~\ref{sec:CW} we briefly remind the main features of the CW/LD scenario; Sec.~\ref{sec:DMPEU} is devoted to the analysis of DM production via freeze-in within the CW/LD model, and 
 in Sec.~\ref{sec:conclusions} we present our conclusions. 
Some relevant decay widths of bulk particles can be found  in App.~\ref{App:decays}.


\input{theoframe}


\input{DMProdEU}

\section{Conclusions}
\label{sec:conclusions}

The Nature of Dark Matter (DM) is one of the most compelling (and long-standing) open problems of our present picture of the Universe: the perfectly working Standard Model (SM) of
Fundamental Interactions is not able to account for it and a plethora of astrophysical and cosmological observations point out that a huge amount of the matter in the Universe
is actually composed of some substance that interacts gravitationally. If gravity is the only way this component of the Universe energy density interacts with the ``standard'' matter
of which we (and the rest of observable bodies in the sky) are made of, how was the observed DM relic abundance produced? One common approach is to consider DM made
of Weakly Interacting Massive Particles (WIMPs), that do interact with ``us'' by means of a new interaction with strength of the order of the weak interactions of the SM.
These particles, however, are supposed to be rather heavy, {\em i.e.}, in the range of GeV to TeV. For WIMPs to be the dominant component of DM in 
the Universe, their abundance could be explained by the so-called ``freeze-out'' mechanism: after an initial phase in which the temperature was high enough so that 
SM and DM particles were in thermal equilibrium, at some point the (heavier) WIMPs keep annihilating into SM particles whereas the opposite reactions become disfavored as the temperature
of the SM bath decrease. Once the DM density becomes so low that it is difficult for two WIMPs to annihilate, their density ``freezes'' and we get a relic abundance that should match what observed today. 
Unfortunately, all experiments devoted to searches (both direct and indirect) for particles with masses in that range
with cross-section as large as weak cross-sections have found no compelling evidence of the existence of such particles. 

An alternative possibility, that we have explored in this paper, was the idea that DM is made by particles with a very small cross-section (sub-weak, or ``feeble'') with SM particles, such that they never reach chemical equilibrium with the SM bath. In this case, the observed DM abundance could have been generated by the so-called ``freeze-in'' mechanism, in which SM particles annihilate into DM ones (``direct freeze-in'')
or into intermediate states that, eventually, decay into DM particles (``sequential freeze-in''). The DM particles produced via such mechanism have been called Feebly Interacting Massive Particles (FIMPs) 
and, typically, can be much lighter than the electro-weak scale, thus explaining their non-observation in direct and indirect detection experiments.

Once a given range of DM masses is selected by choosing a particular mechanism to explain the observed relic abundance, the interesting range of DM cross-section with SM particles is still to be fixed. 
Here is where choosing a specific model in which the interaction strength is dictated by some underlying principle enter into the game. In this paper, we have explored the possibility that the interaction
between DM and SM is, indeed, purely gravitational (as it is suggested by experiments). This was studied previously in the framework of FIMPs (see, e.g., Refs.~\cite{Tang:2017hvq,Garny:2017kha}). 
An interesting way to actually enhance the gravitational interaction, whilst maintaining the feature that DM interacts only gravitationally with the SM, is to embed our model into an extra-dimensional setup. 
Extra-dimensions were proposed in order to explain the so-called ``hierarchy problem'', {\em i.e.}, the huge hierarchy between the electro-weak scale (at which the SM works perfectly) and higher scales at which
new physics should become relevant (for example, the Planck scale at which a quantum theory of gravity would become mandatory). Popular extra-dimensional models are 
Large Extra-Dimensions~\cite{Antoniadis:1990ew, Antoniadis:1997zg, ArkaniHamed:1998rs,Antoniadis:1998ig,ArkaniHamed:1998nn}, Randall-Sundrum models~\cite{Randall:1999ee,Randall:1999vf}
and the more recent Clockwork/Linear Dilaton model~\cite{Antoniadis:2011qw, Cox:2012ee,Giudice:2016yja, Giudice:2017fmj}). In a recent paper \cite{Bernal:2020fvw} we have studied FIMP Dark Matter
in the context of RS models, finding that a significant region of the allowed parameter space is indeed able to reproduce the observed DM relic abundance via the freeze-in mechanism for scalar DM particles with 
mass in the MeV range. In this paper, on the other hand, we have explored the same possibility within the context of the CW/LD model. One of the main difference between the two models is the fact
that in the former Kaluza-Klein--graviton modes arise as isolated resonances (for low KK-numbers) and the model does not differ much from adding a spin-2 resonance to the SM spectrum, whereas in the latter the separation between graviton KK-modes is so tiny that summing over their cumulative effect is unavoidable. As it was the case for RS models, also in the CW/LD model we find a significant region of the 
parameter space in which the observed DM relic abundance can be reproduced with scalar DM in the MeV range, with a reheating temperature that can be as low as 10 GeV. For $\Trh < 100$ GeV , the
typical value of the fundamental scale of gravity, $M_5$, ranges from 10 TeV to 1000 TeV, whereas the first KK-graviton mass $k$ is in the range $k \in [10^2,10^4]$ GeV. 
This rather large allowed region in the model parameter space ($k,M_5$) seems to offer better chances to alleviate the hierarchy problem than in the case of the RS model, where the allowed region 
at low $\Trh$ and low $\Lambda$ can be significantly constrained by LHC Run-III and HL-LHC measurements (with the possible exception of a very peculiar region where the first KK-graviton mass is 
$m_1 \sim 10$ TeV). Exclusion bounds in the CW/LD case from LHC Run-III and the HL-LHC cannot be easily extrapolated from non-resonant searches at the LHC Run-II and will be studied in due time.

Notice that we have assumed an instantaneous decay of the inflaton. In a more realistic picture, in which the instantaneous decay approximation is not used for reheating, 
the SM bath temperature may initially display a temperature much higher than $\Trh$ at onset of the radiation domination~\cite{Giudice:2000ex}.
UV freeze-in is highly sensitive to the dynamics of the thermal bath during the intermediate period between inflation and radiation domination~\cite{Garcia:2017tuj, Bernal:2019mhf} and, therefore, 
it is natural to be analyzed within a specific heating setup (for related studies, see e.g. Refs.~\cite{Chen:2017kvz, Bernal:2018qlk, Bhattacharyya:2018evo, Chowdhury:2018tzw, Kaneta:2019zgw, Banerjee:2019asa, Dutra:2019xet, Dutra:2019nhh, Mahanta:2019sfo, Cosme:2020mck, Bernal:2019mhf, Garcia:2020eof, Bernal:2020bfj, Garcia:2020hyo, Bernal:2020qyu, Anastasopoulos:2020gbu, Criado:2020jkp, Brax:2020gqg}).
It would also be interesting  to analyze  in more detail the cosmology of the CW/LD scenario, in particular the feasibility of complete DM models with such low reheating temperatures as we have found.
These studies are well beyond the scope of this paper, though.

\acknowledgments
NB thanks the theoretical physics department of University of Valencia and the IFIC for their warm hospitality.
This work is also supported by the Spanish MINECO Grants SEV-2014-0398 and FPA2017-84543-P, and by Generalitat Valenciana through the ``plan GenT'' program (CIDEGENT/2018/019) and the grant PROMETEO/2019/083.
NB is partially supported by Universidad Antonio Nariño grants 2018204, 2019101 and 2019248, and the Patrimonio Autónomo - Fondo Nacional de Financiamiento para la Ciencia, la Tecnología y la Innovación Francisco José de Caldas (MinCiencias - Colombia) grant 80740-465-2020.
This project has received funding/support from the European Union's Horizon 2020 research and innovation programme under the Marie Skłodowska-Curie grant agreement No 860881-HIDDeN.
This research made use of IPython~\cite{Perez:2007emg}, Matplotlib~\cite{Hunter:2007ouj} and SciPy~\cite{SciPy}.


\appendix
\input{appendix}

\bibliography{biblio}

\end{document}

%% file: theoframe.tex
\section{Theoretical framework}
\label{sec:CW}

\subsection{A (very) short summary on ClockWork/Linear Dilaton Extra-Dimensions}
\label{sec:clockwork}

In the CW/LD scenario, the following metric is considered  (see Refs.~\cite{Giudice:2016yja,Giudice:2017fmj}):
\be
\label{eq:5dmetric}
ds^2 = G_{MN}^{(5)} dx^M dx^N = e^{4/3 k r_c |y|} \left ( \eta_{\mu \nu} dx^{\mu}dx^{\nu} - r_c^2 \, dy^2 \right ) \, ,
\ee
where $G_{MN}^{(5)}$ is the 5-dimensional metric with signature$(+,-,-,-,-)$ and we use capital Latin indices $M,N$ to run over the 5 dimensions, whereas
Greek indices $\mu,\nu$ run only over the standard 4 dimensions. The extra-dimensional coordinate has been rescaled by the length scale $r_c$, such that $y$ is adimensional.
This particular metric was first proposed in the context of {\em Linear Dilaton} (LD)  models and {\em Little String Theory} (see, e.g., Refs.~\cite{Antoniadis:2011qw,Baryakhtar:2012wj,Cox:2012ee} 
and references therein). 
The metric in eq.~\eqref{eq:5dmetric} implies that the space-time is non-factorizable, as an interval in a 4-dimensional plane scales depending 
on the particular position in the extra-dimension due to the warping factor $\exp (2/3 \, k\, r_c \, |y|)$. Notice that in the limit $k \to 0$ the standard, 
factorizable, flat LED case~\cite{Antoniadis:1990ew,Antoniadis:1997zg,ArkaniHamed:1998rs,Antoniadis:1998ig,ArkaniHamed:1998nn} can be recovered. 
The LED model is, therefore, a particular case of the CW/LD metric.
As it was the case in the Randall-Sundrum model, also in the CW/LD scenario the extra-dimension is compactified on a ${\cal S}_1/{\cal Z}_2$ orbifold (with $r_c$ the compactification radius), 
and two branes are located at the fixed points of the orbifold, $y = 0$ (``IR'' brane) and $y = \pi$ (``UV'' brane).
Standard model fields are located in one of the two branes (usually the IR-brane).
The scale $k$, called the ``clockwork spring'', is the curvature along the 5$^\text{th}$-dimension and it can be much smaller than the Planck scale (remember that, in order to recover the LED case, 
it can even be pushed down to 0). The relation between $M_P$ and the fundamental gravitational scale $M_5$ in the CW/LD model is: 
\be
M_P^2 = \frac{M_5^3}{k} \left ( e^{2 \pi k r_c}  - 1 \right ) \, ,
\ee
and it can be shown that, in order to solve or alleviate the hierarchy problem, $k$ and $r_c$ must satisfy the following relation:
\begin{equation}
\label{eq:krcrelation}
k \, r_c = 10 + \frac{1}{2 \pi} \, \ln \left ( \frac{k}{\rm TeV} \right ) - \frac{3}{2 \pi} \, \ln \left ( \frac{M_5}{ 10 \, {\rm TeV}} \right ) \, .
\end{equation}
In the phenomenological applications of the CW/LD model present in the literature $k$ is typically chosen above the GeV-scale and, therefore, $r_c$ is typically
an extremely small length scale, much smaller than what can be tested in experiments measuring deviations from the $1/r^2$ Newton's law.
Notice that, differently from the case of warped extra-dimensions, where scales are all of the order of the Planck scale ($M_5, k \sim M_P$) or within a few orders of magnitude, 
in the CW/LD scenario, both the fundamental gravitational scale $M_5$ and the mass gap $k$ are much nearer to the electroweak scale $\Lambda_{\rm EW}$ than to the Planck scale, 
as in the LED model. An interesting limit of the CW/LD model is that in which $M_5 = 10$~TeV and $r_c$ saturates the present experimental bound on deviations from the Newton's law, 
$r_c \sim 100$~$\mu$m~\cite{Adelberger:2009zz}. In this ``extreme'' case, eq.~(\ref{eq:krcrelation}) would imply that $k$ could be as small as $ k \sim 2$ eV. In this case, KK-graviton modes 
would be as light as the eV. This particular corner of the parameter space does not differ much from the LED case, but for the important difference that the hierarchy
problem could be solved with just one extra-dimension (for LED models with one extra-dimension, in order to bring $M_5$ down to the TeV scale, an astronomical 
length $r_c$ would be needed). 

The action in 5D is: 
\begin{equation}
{\cal S} = {\cal S}_{\rm gravity} + {\cal S}_{\rm IR} + {\cal S}_{\rm UV}
\end{equation}
where the gravitational part is, in the Jordan frame:
\begin{equation}
\label{eq:5dgravity}
{\cal S}_{\rm gravity} = \frac{M_5^3}{2} \int d^4 x \, \int_0^\pi r_c dy \sqrt{G^{(5)}} \, e^S \, \left [ R^{(5)}  + G_{(5)}^{MN} \partial_M S \partial_N S + 4 k^2  \right ] \, ,
\end{equation}
with $R^{(5)}$ the Ricci scalar and $S$ the (dimensionless) dilaton field, $S = 2 k r_c |y|$. 
We consider for the two brane actions  the following expressions: 
\begin{equation}
\label{eq:IRaction}
{\cal S}_{\rm IR} = \int_0^\pi dy \delta (y) \int d^4 x \, \sqrt{- g_{\rm IR}^{(4)}} \, e^S \left \{ - f_{\rm IR}^4 + {\cal L}_{\rm SM} + {\cal L}_{\rm DM}  \right \}
\end{equation}
and
\begin{equation}
\label{eq:UVaction}
{\cal S}_{\rm UV} =  \int_0^\pi dy \delta (y- \pi) \int d^4 x \, \sqrt{- g_{\rm UV}^{(4)}} \, e^S \left \{ - f_{\rm UV}^4 + \dots \right \} \, ,
\end{equation}
where $f_{\rm IR}, f_{\rm UV}$ are the brane tensions for the two branes, $g_{\rm IR,UV}^{(4)}  = - G^{(5)} / G^{(5)}_{55}$ 
is the determinant of the induced metric on the IR- and UV-brane and the $\delta$-functions at $y=0, \pi$ are adimensional (as $y$ itself).
Throughout the paper, we consider all the SM and DM fields localized on the IR-brane, whereas on the UV-brane we could have
any other physics that is Planck-suppressed (see Ref.~\cite{Kang:2020cxo} for a scenario where all SM fields may be bulk fields). 
We assume that DM particles only interact with the SM particles gravitationally by
considering only DM singlets under the SM gauge group. More complicated DM spectra with several particles and/or particles of different spin will not be studied here. 

Notice that the gravitational action is not in its canonical form. We must first perform a change of frame to cast the action in its ``traditional'' form. This is done by
going to the Einstein frame, changing $G^{(5)}_{MN} \to \exp(-2/3\,S) \tilde G^{(5)}_{MN}$. In the latter frame we get:
\begin{eqnarray}
\label{eq:5dgravity2}
{\cal S}_{\rm gravity} &=& \int d^4 x \, \int_0^\pi r_c dy \sqrt{-\tilde G^{(5)}} \,  \left \{ \frac{M_5^3}{2} 
\left [ R^{(5)}  - \frac{1}{3} \tilde G_{(5)}^{MN} \partial_M S \partial_N S + 4 e^{- \frac{2}{3} S} k^2  \right ] \right \}
\nonumber \\
&+& \int d^4 x \, \int_0^\pi dy \sqrt{- \tilde g^{(4)}} \, e^{-\frac{S}{3}} 
\left \{ \delta (y ) \left [ - f^4_{\rm IR} +  {\cal L}_{\rm SM} + {\cal L}_{\rm DM} \right ] - \delta (y - \pi) f^4_{\rm UV} \right \},\quad
\end{eqnarray}
where now the gravitational action is just the Ricci scalar. 
The kinetic term of the dilaton field tells us that the physical field must be rescaled as $\hat S = \left ( M_5^{3/2}/\sqrt{3}\right ) \, S$, 
with $\hat S$ a canonical 5-dimensional scalar field (with dimension $[\hat S] = 3/2$). 
Notice that, in the Einstein frame, the brane action terms still have an exponential dependence $e^{-S/3}$ from the dilaton field. 
This action has a shift symmetry $S \to S + {\rm const}$ in the limit $k \to 0$, that makes a small value of $k$ with respect to $M_5$ ``technically
natural'' in the 't Hooft sense. Using the action above in the Einstein frame, it can be shown that the metric in eq.~(\ref{eq:5dmetric}) can be
recovered as a classical background if the brane tensions are chosen as: 
\begin{equation}
\label{eq:kfinetuning}
f^4_{\rm IR} = - f^4_{\rm UV} = - 4 k \, M_5^3 \, .
\end{equation}

Notice that, in a pure 4-dimensional scenario, the gravitational interactions would be enormously suppressed by powers of the Planck mass, 
while in an  extra-dimensional one the gravitational interaction is actually enhanced. 
Expanding the metric at first order around its static solution, we have: 
\be
\label{eq:metricexpansion}
G^{(5)}_{MN} =  e^{2/3  S} \left ( \eta_{MN} + \frac{2}{M_5^{3/2}} h_{MN} \right ) \, .
\ee
The 4-dimensional component%
\footnote{
In the decomposition of $h_{MN}$ two other fields are present: the graviphoton $h_{\mu5}$ and the graviscalar $h_{55}$ (also called $\Phi$ following the 
notation of Ref.~\cite{Cox:2012ee}). 
The graviscalars KK-tower is absent from the low-energy spectrum, as they are eaten by the KK-tower of graviphotons to get a mass. 
These are, in turn, eaten by the KK-gravitons to get a mass. The resulting massive KK-gravitons have, thus, five degrees of freedom. 
The only degrees of freedom still present at low-energy are the graviphoton zero-mode (that does not couple with the energy-momentum tensor in the weak gravitational field limit~\cite{Giudice:1998ck}),
and the graviscalar zero-mode. The latter will generically mix with the radion needed to stabilize the extra-dimension size.
}
of the 5-dimensional field $h_{MN}$ can be expanded in a Kaluza-Klein tower of 4-dimensional fields as follows:
\begin{equation}
h_{\mu\nu} (x,y) =  \frac{1}{\sqrt{\pi r_c}} \sum_{n = 0}^\infty h^n_{\mu\nu} (x) \, \chi_n (y) \, ,
\end{equation}
so that $h^n_{\mu\nu}(x)$ are canonical 4-dimensional scalar fields (with dimension 1) and
the adimensional $\chi_n (y)$ eigenfunctions depict the profile of the $h^n_{\mu\nu} (x)$ KK-modes in the extra-dimension.
They can be computed by solving the equation of motion in the extra-dimension of the fields: 
\begin{equation}
\left [ \partial_y^2 - k^2 r_c^2 + m_n^2 r_c^2 \right ] e^{k r_c |y|} \, \chi_n (y) = 0
\end{equation}
with Neumann boundary conditions $\partial_y \chi_n (y) = 0$ at $y = 0$ and $\pi$. Normalizing the eigenmodes such that the KK-modes have
canonical kinetic terms in 4-dimensions, we get: 
\begin{equation}
\left \{
\begin{array}{lll}
\chi_0 (y) & = & \sqrt{\frac{\pi k r_c}{e^{2 \pi k r_c} - 1}} \, , \\
&& \\
\chi_n (y) & = & \frac{n}{m_n r_c} e^{- k r_c |y|} \left ( \frac{k r_c}{n} \sin n |y| + \cos n  |y| \right ) \, ,
\end{array}
\right .
\end{equation}
with masses
\begin{equation}
\label{eq:KKgravmasses}
m_0^2 = 0 \, ; \qquad m_n^2 = k^2 + \frac{n^2}{r_c^2} \, .
\end{equation}

At the IR-brane one gets: 
\be
\mathcal{L} = -\frac{1}{M_5^{3/2}} T^{\mu \nu}(x) h_{\mu \nu}(x,y=0) = - \sum_{n=0} \frac{1}{\Lambda_n}  h^{n}_{\mu \nu} (x) \, T^{\mu \nu}(x)\, ,
\ee
where 
\begin{equation}
\left \{
\begin{array}{lll}
\frac{1}{\Lambda_0} &=& \frac{1}{M_P} \, ,\\
&& \\
\frac{1}{\Lambda_n} &=&  \frac{1}{\sqrt{\pi r_c M_5^3 }}  \left ( 1 + \frac{k^2 r_c^2}{n^2} \right )^{-1/2} 
=  \frac{1}{\sqrt{ \pi r_c M_5^3}} \left ( 1 - \frac{k^2}{m_n^2} \right )^{1/2}  \, \qquad n \neq 0 \, ,
\end{array}
\right .
\label{Lambda_graviton}
\end{equation}
and the explicit form of the energy-momentum tensor for a scalar field $\Phi$ is:
\begin{equation}
T_{\mu \nu}^\Phi = (\partial_\mu \Phi)^\dagger (\partial_\nu \Phi) + (\partial_\nu \Phi)^\dagger (\partial_\mu \Phi) - \eta_{\mu \nu} \left\lbrace (\partial_\rho \Phi)^\dagger (\partial\rho \Phi) - m_\Phi^2 \Phi \Phi^\dagger \right\rbrace \, .
\end{equation}
It can be seen that the coupling between KK-graviton modes with $n \neq 0$ is suppressed by the effective scale $\Lambda_n$
and not by the Planck scale, differently from the LED case and similarly to the Randall-Sundrum one. In the latter model, however, $\Lambda$ is a universal scale, whereas in the CW/LD model
it depends on the KK-number $n$.

\subsection{Stabilization mechanism}
\label{sec:rad}

Stabilization of the radius of the extra-dimension $r_c$ is an issue in all extra-dimensional models. The standard lore (see, e.g., Refs.~\cite{Appelquist:1982zs,Appelquist:1983vs}) is that 
bosonic quantum loops have a net effect on the boundaries of an extra-dimension such that the latter shrinks to a point~\cite{deWit:1988xki}. This feature can be compensated by 
fermionic quantum loops that give a positive contribution to the Casimir energy and, thus, would expand the size of the extra-dimension. Within supersymmetric formulation of extra-dimensional models, 
these two effects may compensate each other and stabilize the radius of the extra-dimension (see, e.g., Ref.~\cite{Ponton:2001hq}). Supersymmetry may also be useful to protect the CW/LD background
metric by fluctuations of the 5-dimensional cosmological constant and of the brane tensions in eq.~(\ref{eq:kfinetuning}). Non-supersymmetric clockwork implementations do exist, though~\cite{Teresi:2018eai}.

Differently from the Randall-Sundrum case (where a new scalar field must be added to the Lagrangian), in the CW/LD scenario we can use the bulk dilaton field $S$ to stabilize the compactification radius. 
Adding to the action in eq.~(\ref{eq:IRaction}) some ad-hoc localized brane interactions for $S$ at $y = 0, \pi$, we could fix the value of $S$ at the UV-brane, $ S(\pi)$,
such that $S(\pi) = 2 \pi k \,  r_c $~\cite{Giudice:2016yja}:
\begin{equation}
\label{eq:localizedpotentials}
\left \{
\begin{array}{l}
{\cal S}_{\rm IR} = \int d^4 x \, \sqrt{- g_{\rm IR}^{(4)}} \, e^S \left \{ - f_{\rm IR}^4 + \frac{\mu_{\rm IR}}{2} \left ( \hat S - \hat S_{\rm IR} \right )^2 + {\cal L}_{\rm SM} + {\cal L}_{\rm DM}  \right \} \, , \\
\\
{\cal S}_{\rm UV} = \int d^4 x \, \sqrt{- g_{\rm UV}^{(4)}} \, e^S \left \{ - f_{\rm UV}^4 + \frac{\mu_{\rm UV}}{2} \left ( \hat S - \hat S_{\rm UV} \right )^2 + \dots \right \} \, ,
\end{array}
\right .
\end{equation}
with $\mu_{\rm IR}$ and $\mu_{\rm UV}$ two parameters with the dimension of a mass and $\hat S$ the rescaled dilaton field. 

The physical scalar spectrum of the model includes the zero-mode of the dilaton $\hat S$, its KK-modes tower and the zero-mode of the graviscalar $\Phi$. 
We introduce quantum fluctuations over the background values of $\hat S (x,y) = \hat S_0(y) + \varphi (x,y)$, where $\hat S_0 (y) = 2/\sqrt{3}\,M_5^{3/2} \, k\ r_c\, |y|$, 
and of the metric in eq.~\eqref{eq:metricexpansion}, in order to derive the equation of motions for the two scalar degrees of freedom, $\varphi$ and $\Phi$.
After deriving the Einstein equations for $\varphi$ and $\Phi$, and imposing the junction conditions at the boundaries, we find:%
\begin{equation}
\left [ \Box + \frac{1}{r_c^2} \frac{d^2}{dy^2} - k^2 \right ] e^{k r_c y} \left ( 
\begin{array}{l}
\Phi (x, y) \\
\varphi (x,y) 
\end{array}
\right ) = 0 \, .
\end{equation}
Notice that the physical combination of $\Phi$ and $\varphi$ fields with a canonical kinetic term was identified in Ref.~\cite{Kofman:2004tk}, and it corresponds to the definition:
\begin{equation}
\label{footnote:2}
v (x,y) = \sqrt{6} e^{k r_c y} \left [ \Phi (x,y) - \frac{1}{\sqrt{3}} \varphi(x,y) \right ] \, .
\end{equation}

We can expand a generic scalar field $\phi$ (representing either $\Phi$, $\varphi$ or $v$) over a 4-dimensional plane-waves basis, 
\begin{equation}
\phi (x,y) =  \frac{1}{\sqrt{\pi r_c}} \sum_n Q_n (x) \phi_n (y)  \, 
\end{equation}
where the 4-dimensional fields $Q_n$ satisfy the Klein-Gordon equation $\left [ \Box - m_{\phi_n}^2 \right ] Q_n = 0$ and the adimensional eigenfunctions
(for example, in $v$) are given by: 
\begin{equation}
\label{eq:CWLIDeigenmodes}
v_n (y) = N_n e^{- k r_c y} \left [ \sin (\beta_n r_c \, y) + \omega_n \cos (\beta_n r_c \, y) \right ]  \, ,
\end{equation}
with $N_n$ a normalization factor, $\beta_n^2 = m_{v_n}^2 - k^2$, and
\begin{equation}
\omega_n = - \frac{3 \beta_n \mu_{\rm IR}}{2 (k^2 + \beta_n^2) + k \mu_{\rm IR}} \, ,
\end{equation}
using the boundary condition at the UV-brane to fix one of the constants.
In the so-called {\em rigid limit}, $\mu_{\rm IR} \to \infty$ and $\mu_{\rm UV} \to \infty$, the scalar spectrum (first obtained in Ref.~\cite{Kofman:2004tk}) is given by: 
\begin{equation}
\label{eq:KKdilatonmasses}
\left \{ 
\begin{array}{l}
m_r^2 \equiv m_{v_0}^2 = \frac{8}{9} k^2 \, ,\\
\\
m_{v_n}^2 = k^2 + \frac{n^2}{r_c^2} \qquad ({\rm for} \, n \geq 1) \, , 
\end{array}
\right .
\end{equation}
where we have identified the radion as the lightest state. 
Out of the rigid limit, the spectrum can be obtained expanding in inverse powers of $\mu_{\rm UV}$ and $\mu_{\rm IR}$, 
introducing the adimensional parameters $\epsilon_{\rm IR, UV} = 2 k / \mu_{\rm IR, UV}$. At first order in the $\epsilon$-parameters (see Ref.~\cite{Cox:2012ee}), 
\begin{equation}
\left \{ 
\begin{array}{l}
m_r^2 \equiv m_{v_0}^2 = \frac{8}{9} k^2 \left ( 1 - \frac{2 \epsilon_{\rm IR}}{9} \right ) + {\cal O} (\epsilon_\text{UV}^2,\epsilon_\text{IR}^2) \, ,\\
\\
m_{v_n}^2 = k^2 + \frac{n^2}{r_c^2} \left [ 1 - \frac{6 (n^2 + k^2 r_c^2) (\epsilon_{\rm UV} + \epsilon_{\rm IR})}{9 n^2 \pi k r_c + \pi k^3 r_c^3} \right ] + {\cal O} (\epsilon_\text{UV}^2,\epsilon_\text{IR}^2) \, .
\end{array}
\right .
\end{equation}
There are no massless states for non-vanishing $\mu$'s ({\em e.}, when the extra-dimension is stabilized). In the unstabilized regime,
$\mu_{\rm UV}, \mu_{\rm IR} \to 0$, the graviscalar and lowest-lying dilaton mode decouple 
(as the eigenmodes spectrum in eq.~(\ref{eq:CWLIDeigenmodes}) for $\omega_n \to 0$ starts at $n=1$)
and we expect two massless modes (see, again, Ref.~\cite{Cox:2012ee}). 

The interactions of the radion and of the dilaton KK-tower with SM fields arises~\cite{Cox:2012ee} from the term: 
\begin{equation}
\int d^4 x \sqrt{- g^{(4)}} \, e^{-S/3} \left [ {\cal L}_{\rm SM} + {\cal L}_{\rm DM}\right ] \, .
\end{equation}
Whilst in the Randall-Sundrum model the brane action term couples to gravity minimally~\cite{Goldberger:1999un}, {\em i.e.}, through the $\sqrt{-g^{(4)}}$ coefficient, only, in the CW/LD model 
when going from the Jordan to the Einstein frame a dilaton exponential coupling with the brane-localized SM Lagrangian is still present. 
The action, after KK-decomposition, can be expanded at first order in quantum fluctuations over the metric and of the dilaton field: 
\begin{eqnarray}
{\cal S}_{\rm int} &=& - \frac{1}{2} \frac{1}{\sqrt{\pi r_c M_5^3}} \sum_n \Phi_n (0 ) \int d^4 x \left [ g_0^{(4)} \right ]^{\mu\nu} \left [ T^{\rm SM}_{\mu\nu} + T^{\rm DM}_{\mu\nu} \right ] Q_n \nonumber \\
&-&  \frac{1}{3} \sqrt{\frac{3}{\pi r_c M_5^3}} \sum_n \varphi_n (0 ) \int d^4 x  \left [ {\cal L}_{\rm SM} + {\cal L}_{\rm DM}  \right ] Q_n \, ,
\end{eqnarray}
where, as expected, scalar fluctuations of the metric ($\Phi$) and of the dilaton field ($\varphi$) couple with SM fields through two terms: 
the usual energy-momentum trace and a new term arising from the exponential 
coupling of the dilaton with the brane Lagrangian in the Einstein frame. Notice, however, that the physical field $v_n$ in eq.~(\ref{eq:CWLIDeigenmodes}) 
will couple with both terms at the same time.

At first order in $\epsilon_{\rm UV, IR}$, we get for the couplings with the energy-momentum tensor: 
\begin{equation}
\left \{
\begin{array}{lll}
\frac{1}{\Lambda_{\Phi_0}} \equiv \frac{\Phi_0(0)}{2 \sqrt{\pi r_c M_5^3}} &=& \frac{1}{6} \sqrt{\frac{k}{M_5^3} } \left ( 1 + \frac{4}{9} \epsilon_{\rm UV} \right ) + {\cal O}(\epsilon_\text{UV}^2,\epsilon_\text{IR}^2) \, , \\
&&\\
\frac{1}{\Lambda_{\Phi_n}} \equiv \frac{\Phi_n(0)}{2 \sqrt{\pi r_c M_5^3}} &=& \frac{2 k r_c n}{\sqrt{3 \pi r_c M_5^3} }    \left ( n^2 + k^2 r_c^2 \right )^{-1/2}   \left ( 9 n^2 + k^2 r_c^2 \right )^{-1/2} 
(1 - \epsilon_{\rm UV}) 
+ {\cal O}(\epsilon_\text{UV}^2,\epsilon_\text{IR}^2) \\
&=& \frac{2 }{\sqrt{27 \pi r_c M_5^3} } \, \frac{k}{m_{v_n}} \left [  \frac{ \left ( 1 - k^2 / m_{v_n}^2  \right ) }{ \left ( 1 - \frac{8}{9} k^2 / m_{v_n}^2 \right ) } \right ]^{1/2}
 (1 - \epsilon_{\rm UV}) + {\cal O}(\epsilon_\text{UV}^2,\epsilon_\text{IR}^2)
\end{array}
\right .
\label{Lambda_radion}
\end{equation}
and for the couplings with the Lagrangian:
\begin{equation}
\left \{
\begin{array}{l}
\frac{1}{\Lambda_{\varphi_0}} \equiv \frac{\varphi_0 (0)}{\sqrt{3 \pi r_c M_5^3}} = \frac{2}{27} \sqrt{ \frac{k}{M_5^3} } \epsilon_{\rm UV} + {\cal O}(\epsilon_\text{UV}^2,\epsilon_\text{IR}^2) \, , \\
\\
\frac{1}{\Lambda_{\varphi_n}} \equiv \frac{\varphi_n (0)}{\sqrt{3 \pi r_c M_5^3}} = \frac{n}{k \sqrt{3 \pi M_5^3 r^3_c} }  
\left [ \frac{ \left ( n^2 + k^2 r_c^2 \right ) }{ \left ( 9 n^2 + k^2 r_c^2 \right ) } \right ]^{1/2} \epsilon_{\rm UV} + {\cal O}(\epsilon_\text{UV}^2,\epsilon_\text{IR}^2) \, . \\
\end{array}
\right .
\end{equation}
In the rigid limit ($\mu_{\rm UV, IR} \to \infty$) the coupling of dilaton modes with the SM Lagrangian vanishes
($1/\Lambda_{\varphi_0}, 1/\Lambda_{\varphi_n} \to 0$). 
In the rest of the paper, we will work in this limit when computing the effects of radion and KK-dilatons in the freeze-in mechanism. 
A complete study of the impact of scalar perturbations to the DM phenomenology outside of this limit is beyond the scope of this paper.

Eventually, we are not taking into account possible terms that may arise coupling scalar fields (such the Higgs boson, or the DM particle considered here) and the Ricci scalar, 
that are usually not forbidden by any symmetry of the theory: 
\begin{equation}
\label{eq:nonminimalHiggsmixing}
\Delta {\cal S}_{\rm IR} = \int d^4 x \sqrt{- g^{(4)}} e^{S/3} \xi R H^\dagger H \, .
\end{equation}
Such terms induce an additional kinetic mixing between the graviscalar $\Phi_0$, the lowest-lying dilaton $\varphi_0$ and the Higgs and, therefore, 
additional couplings with the SM fields. We will neglect this non-minimal coupling in the rest of the paper, taking $\xi = 0$. 

Summarizing, in the rigid limit and in the absence of a mixing between the Higgs and the other scalar fields, the scalar perturbation interaction 
Lagrangian with SM and DM particles at first order is: 
\begin{equation}
{\cal L}^{\rm SM}_v = \sum_{n=0}^\infty \frac{1}{\Lambda_{\Phi_n}}  \left [
T_{\rm SM}+ \frac{\alpha_\text{EM} \, C_{\rm EM}}{8\pi} F_{\mu\nu} F^{\mu\nu} + \frac{\alpha_s C_{3}}{8 \pi} \sum_a F^a_{\mu\nu} F^{a\mu\nu}
\right ] \, v_n \, , 
\label{radion_lag}
\end{equation}
where $r = v_0$ is the radion field and $v_n$ for $n \geq 1$ is the dilaton KK-tower, and $T_{\rm SM}$ is the trace of the SM energy-momentum tensor. 
In eq.~(\ref{radion_lag}), we have~\cite{Blum:2014jca}: 
\be
\left \{
\begin{array}{lll}
C_3 &=& b_\text{IR}^{(3)} - b_\text{UV}^{(3)} + \frac{1}{2}\sum_q F_{1/2}(x_q) \, , \\
&&\\
C_\text{EM} &=& b_\text{IR}^\text{(EM)} - b_\text{UV}^\text{(EM)} + F_1(x_W)  - \sum_q N_cQ_{q}^2F_{1/2}(x_q) \, ,
\end{array}
\right .
\ee
with $x_q = 4m_q/m_r$ and $x_W = 4m_W/m_r$. The values of the one-loop $\beta$-function coefficients $b$ are 
$b_\text{IR}^{(\rm EM)} - b_\text{UV}^{(\rm EM)} = 11/3$ and $b_\text{IR}^{(3)} - b_\text{UV}^{(3)} = -11 + 2n/3$, 
where $n$ is the number of quarks whose mass is smaller than $m_r/2$.
The explicit form of $F_{1/2}$  and $F_1$ is given by:
\be
\left \{
\begin{array}{lll}
F_{1/2}(x) = 2x[1 + (1-x)f(x)] , \\
&&\\
F_{1}(x) = 2 + 3x + 3x(2-x)f(x) , 
\end{array}
\right .
\ee
with
\be
f(x) = \left \{
\begin{array}{lll}
\left[\arcsin(1/\sqrt{x})\right]^2 \hphantom{} \hphantom{} \hphantom{} \hphantom{} \hphantom{} &x>1 , \\
&&\\
-\frac{1}{4}\left[\log\left( \frac{1 + \sqrt{x-1}}{1 - \sqrt{x-1}} \right) - i\pi \right]^2 &x<1 .
\end{array}
\right .
\ee

%% file: DMProdEU.tex
\section{Dark Matter Production}
\label{sec:DMPEU}

The evolution of the DM, radion/KK-dilatons and KK-gravitons number densities ($n$, $n_v$ and $n_G$, respectively) can be followed by the system of coupled Boltzmann equations
\begin{eqnarray}\label{eq:cosmo1}
	\frac{dn}{dt}+3\,H\,n&=&-\gamma_\text{DM$\to$SM}\left[\left(\frac{n}{n^\text{eq}}\right)^2-1\right]+\gamma^d_{G_n\to\text{DM}}\left[\frac{n_G}{n_G^\text{eq}}-\left(\frac{n}{n^\text{eq}}\right)^2\right]\nonumber\\
	&&+\gamma^d_{v_n\to\text{DM}}\left[\frac{n_v}{n_v^\text{eq}}-\left(\frac{n}{n^\text{eq}}\right)^2\right],\\
	\frac{dn_v}{dt}+3\,H\,n_v&=&-\gamma_{v_n\to\text{SM}}\left[\left(\frac{n_v}{n_v^\text{eq}}\right)^2-1\right]-\gamma^d_{v_n\to\text{DM}}\left[\frac{n_v}{n_v^\text{eq}}-\left(\frac{n}{n^\text{eq}}\right)^2\right]\nonumber\\
	&&-\gamma^d_{v_n\to\text{SM}}\left[\frac{n_v}{n_v^\text{eq}}-1\right],\label{eq:cosmo3}\\
	\frac{dn_G}{dt}+3\,H\,n_G&=&-\gamma_{G_n\to\text{SM}}\left[\left(\frac{n_G}{n_G^\text{eq}}\right)^2-1\right]-\gamma^d_{G_n\to\text{DM}}\left[\frac{n_G}{n_G^\text{eq}}-\left(\frac{n}{n^\text{eq}}\right)^2\right]\nonumber\\
	&&-\gamma^d_{G_n\to\text{SM}}\left[\frac{n_G}{n_G^\text{eq}}-1\right],\label{eq:cosmo2}
\end{eqnarray}
where $H$ corresponds to the Hubble expansion rate and $n_i^\text{eq}$ are the equilibrium number densities of the species $i$.
Moreover, $\gamma_{\alpha\to\beta}$ and $\gamma^d_{\alpha\to\beta}$ are the interaction rate densities for the 2-to-2 annihilations and 1-to-2 decays of $\alpha$ into $\beta$, respectively.
Remind that eqs.~\eqref{eq:cosmo3} and~\eqref{eq:cosmo2} refer to a generic KK-mode of the radion/KK-dilaton or KK-graviton towers. Therefore, if we may kinematically produce $N_v$ and $N_G$
radion/KK-dilaton and KK-graviton modes, respectively, the complete Boltzmann equations system consists of $N_v + N_G + 1$ coupled equations.

If DM never reaches chemical equilibrium with the SM plasma (so that the freeze-out mechanism is not viable), it could have been created in the early Universe via the freeze-in paradigm.
There are two possible implementations of the freeze-in mechanism:  the DM abundance could have been generated directly by the annihilation of SM particles via an $s$-channel exchange of KK-gravitons or radions, 
${\rm SM\,SM} \to {\rm DM\,DM}$, through the so-called \textit{direct freeze-in}; or, alternatively, the DM abundance could have also been generated by decays of KK-gravitons or radions, 
previously produced by annihilations, ${\rm SM\,SM} \to G_n G_m$, $G_n v_m$, $v_n v_m$,
or inverse decays of SM particles, 
${\rm SM\,SM} \to G_n$, $v_n $,  via direct freeze-in
(this scenario has been dubbed \textit{sequential freeze-in}~\cite{Hambye:2019dwd, Belanger:2020npe}).

Few comments are in order.
The DM abundance could also be set entirely in the hidden sector by the dark freeze-out of 4-to-2 interactions, where four DM particles annihilate into two of them~\cite{Carlson:1992fn, Bernal:2015xba, Bernal:2017mqb, Bernal:2020gzm}.
However, such a possibility is sub-dominant due to a strong suppression by higher orders of the scale $M_5$.
Moreover, even if it is possible to populate the relic abundance via the exchange of a massless spin-2 graviton~\cite{Garny:2015sjg, Tang:2017hvq, Garny:2017kha, Bernal:2018qlk}, this channel is suppressed by $M_P^4$, therefore being typically subdominant.

The two main mechanisms previously mentioned, {\em i.e.}, direct and sequential freeze-in, will be described in detail in the following subsections.

\subsection{Direct Freeze-in}
\label{sec:directFI}

Direct freeze-in occurs when DM is mainly produced through the annihilation of SM particles via an $s$-channel exchange of KK-gravitons or radions.
In that case, eqs.~\eqref{eq:cosmo1}-\eqref{eq:cosmo2} can be rewritten in terms of the dimensional DM yield $Y\equiv n/\mathfrak{s}$, with $\mathfrak{s}(T)=\frac{2\pi^2}{45}\gss(T)\,T^3$ the SM entropy density defined in terms of the effective number of relativistic degrees of freedom $\gss$~\cite{Drees:2015exa}, as
\begin{equation}\label{eq:dYdTdirect}
	\frac{dY}{dT} \simeq \frac{\gamma_\text{DM$\to$SM}}{H\,\mathfrak{s}\,T}\left[\left(\frac{Y}{Y^\text{eq}}\right)^2-1\right] \simeq -\frac{\gamma_\text{DM$\to$SM}}{H\,\mathfrak{s}\,T}.
\end{equation}
In the last step the fact that DM is always out of chemical equilibrium was used, {\em i.e.}, $Y\ll Y^\text{eq}$.
Equation~\eqref{eq:dYdTdirect} admits an approximate analytical solution given by
\begin{equation}\label{eq:Ydirect}
	Y(T)\simeq\frac{135}{2\pi^3\,\gss}\sqrt{\frac{10}{\gs}}\,M_P\int_{\Trh}^T\frac{\gamma_\text{DM$\to$SM}}{T^6}\,dT\,,
\end{equation}
using the fact that $H^2=\frac{\rho_\text{SM}}{3M_P^2}$, with the SM energy density $\rho_\text{SM}(T)=\frac{\pi^2}{30}\gs(T)\,T^4$ and $\gs(T)$ being the effective number of relativistic degrees of freedom for the SM radiation~\cite{Drees:2015exa}.
Within the approximation of a sudden decay of the inflaton, the maximal temperature reached by the SM thermal bath is the {\em reheating temperature}, $\Trh$.

The interaction rate density $\gamma_\text{DM$\to$SM}$ can be computed from the total DM annihilation cross-section into SM states $\sigma_\text{DM$\to$SM}$ (i.e., the sum over the 2-to-2 DM annihilation cross-sections into all possible SM final states) by the use of eq.~\eqref{eq:gamma_sin_approx}.
In the limit where the DM and SM masses are neglected, $\sigma_\text{DM$\to$SM}$ can be expressed as
\begin{equation}\label{virtual_exchange}
	\sigma_\text{DM$\to$SM}(s) \simeq \frac{49 \, s^3}{1440 \pi} \left|\sum_{n=1}^\infty\frac{1}{\Lambda_n^2}\frac{1}{s-m_n^2+i\,m_n\Gamma_n}\right|^2 + \frac{s^3}{8\pi}\left|\sum_{n=0}^\infty\frac{1}{\Lambda_{\Phi_n}^2}\frac{1}{s-m_{v_n}^2+i\,m_{v_n}\Gamma_{v_n}}\right|^2,
\end{equation}
where the two terms correspond to the exchange of KK-gravitons and radion/KK-dilatons, respectively.
We want to note that in order to have an analytical estimation, in eq.~\eqref{virtual_exchange} the interference terms between KK-gravitons and radion/KK-dilatons have been ignored. However, they are taken into account in the full numerical analysis.

\begin{figure}[t]
	\begin{center}
		\includegraphics[height=0.32\textwidth]{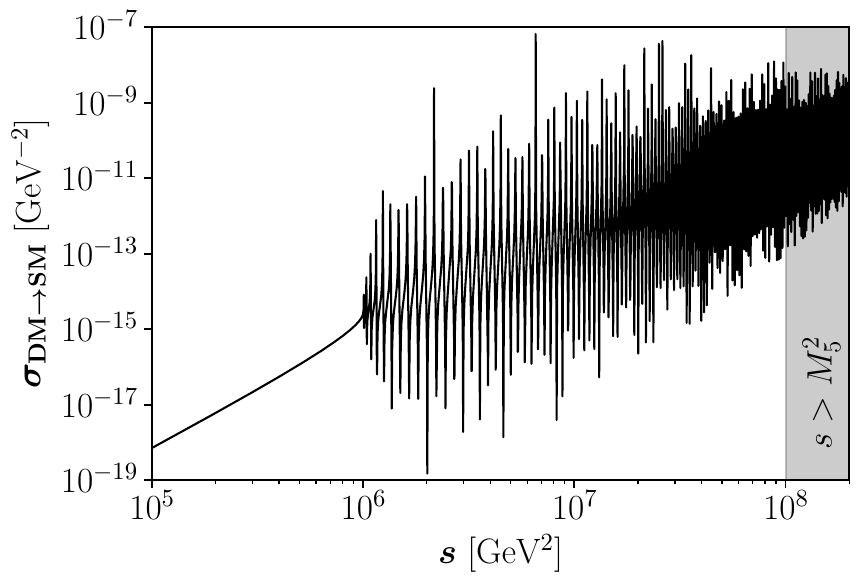}
		\includegraphics[height=0.32\textwidth]{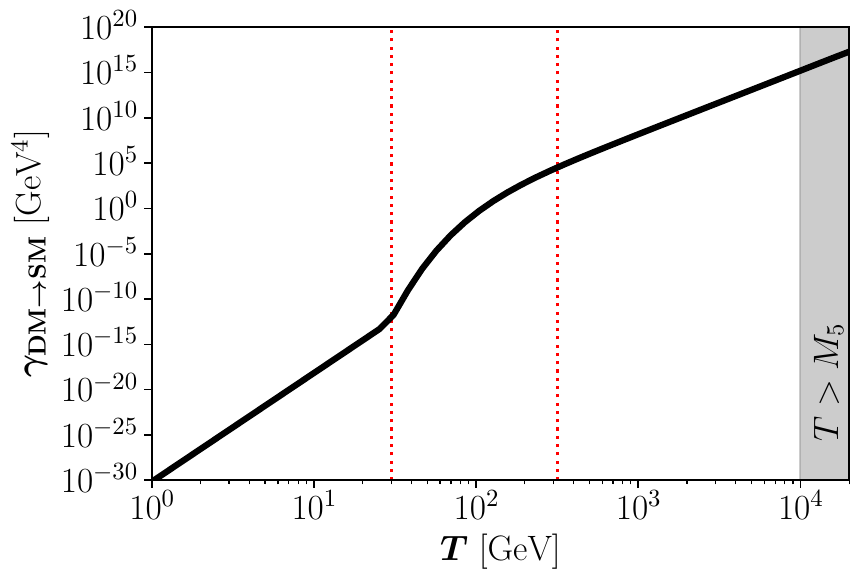}
		\caption{Black solid lines represent the DM annihilation cross-section (left panel) and interaction rate density (right panel) for $k=10^3$~GeV, $M_5=10^4$~GeV and $m_\text{DM}=1$~MeV.
		Red-dotted vertical lines in the right panel depict the different regimes in eq.~\eqref{eq:gammaann}.
		The gray-shaded regions on the right of both panels are beyond the EFT approach.
		} \label{fig:cross}
	\end{center}
\end{figure}
The left panel of Fig.~\ref{fig:cross} shows the DM annihilation cross section $\sigma_\text{DM$\to$SM}$ for $k=10^3$~GeV, $M_5=10^4$~GeV and $m_\text{DM}=1$~MeV.
Notice, however, that this cross-section is largely independent of the DM mass as long as $m_\text{DM}\ll \sqrt{s}$. We have found that this process is dominated by the virtual KK-graviton exchange. Indeed, the difference between the couplings of gravitons and dilatons (eqs.~\eqref{Lambda_graviton} and~\eqref{Lambda_radion}, respectively) makes negligible the dilaton contributions after the first resonance.
Eventually, the gray-shaded region corresponding to $s>M_5^2$ is beyond our EFT approach, being the center-of-mass energy of the process larger than the effective scale of the theory.

The right panel of Fig.~\ref{fig:cross} shows the DM interaction rate density $\gamma_\text{DM$\to$SM}$ entering in eq.~\eqref{eq:Ydirect}, for a particular point of the parameter space, 
$k=10^3$~GeV, $M_5=10^4$~GeV and $m_\text{DM}=1$~MeV. The latter is derived from the cross section $\sigma_\text{DM$\to$SM}$ and can be approximated by the following expressions: 
\begin{equation}\label{eq:gammaann}
	\gamma_\text{DM$\to$SM}(T)\simeq
	\begin{cases}
		1.7\times 10^{-1}\,\mathcal{C} (k \, r_c)\,\frac{T^{12}}{M_5^6\,k^2} &\qquad\text{ for }T\ll m_1/2\,,\\[8pt]
		1.2\times 10^{-3}\,\mathcal{C} (k \, r_c)\,\frac{r_c\,k^7\,T}{M_5^3}\,K_1\left(\frac{m_1}{T}\right) &\qquad\text{ for }T\simeq m_1/2\,,\\[8pt]
		2.8\times 10^{-3}\,\mathcal{C} (k \, r_c)\,\frac{r_c\,k\,T^7}{M_5^3} &\qquad\text{ for }T\gg m_1/2\,,
	\end{cases}
\end{equation}
where
\begin{equation}
\label{eq:Cfunction}
	\mathcal{C} (k \, r_c) \equiv\left[\frac{\sinh(2\pi\,k\,r_c)-2\pi\,k\,r_c}{\sinh^2(\pi\,k\,r_c)}\right]^2
\end{equation}
comes from the sum over all the KK modes, with $\mathcal{C} (k \, r_c)\simeq 4$ when $k\,r_c>1$.

At low temperatures ($T\ll m_1/2$) all the mediators are very heavy and decouple from the low-energy theory; the rate presents a strong temperature dependence, $\gamma \propto T^{12}$. In this regime,  it is possible to perform analytically the sum over all the KK modes that contribute  to the cross section in eq.~\eqref{virtual_exchange},  in the limit $s \ll m_1^2$ (see App.~\ref{app:kksum}).
When the temperature approaches the mass of the first KK-graviton (and first radion), its resonant exchange dominates and $\gamma\propto T\,K_1(m_1/T)$.
Nearby resonances are very close in mass and, therefore, for $T\gg m_1/2$ several KK-gravitons give sizable contributions, with a constructive interference giving the observed $\gamma\propto T^7$ behavior.
As for the left panel, the gray-shaded region corresponding to $T>M_5$ is beyond our EFT approach. 


The approximations of eq.~\eqref{eq:gammaann} allow to analytically solve eq.~\eqref{eq:Ydirect}.
Its asymptotic solution $Y_0$ for $T\ll\Trh$ is
\begin{equation} 
	Y_0\simeq
	\begin{cases}
		\frac{5.3\times 10^{-2}}{\gss}\sqrt{\frac{10}{\gs}}\,\mathcal{C}\,\frac{M_P\,\Trh^7}{M_5^6\,k^2} &\qquad\text{ for }\Trh\ll m_1/2\,,\\[8pt]
		\frac{1.2\times 10^{-4}}{\gss}\sqrt{\frac{10}{\gs}}\mathcal{C}\frac{M_P\,r_c\,k^{7/2}}{M_5^3}\frac{12k^2+10\sqrt{3}k\,\Trh+15\Trh^2}{\Trh^{5/2}}\,e^{-\frac{\sqrt{3}k}{\Trh}} &\qquad\text{ for }\Trh\simeq m_1/2\,,\\[8pt]
		\frac{3.0\times 10^{-3}}{\gss}\sqrt{\frac{10}{\gs}}\,\mathcal{C}\,\frac{M_P\,r_c\,k\,\Trh^2}{M_5^3} &\qquad\text{ for }\Trh\gg m_1/2\,.
	\end{cases}
\end{equation}
The final DM yield has a strong dependence on $\Trh$, characteristic of the UV freeze-in production mechanism.

Now, let us emphasize that for the previous analysis to be valid, the DM has to be out of chemical equilibrium with the SM bath.
This is guaranteed as long as $\gamma_\text{DM$\to$SM}\ll H\,n_\text{eq}$, which translates into
\begin{equation} 
	\Trh\ll
	\begin{cases}
		\left(\frac{\gs}{10}\right)^{1/14}\left(\frac{M_5^6 \, k^2}{\mathcal{C} \, M_P}\right)^{1/7} \,  &\qquad\text{ for }\Trh\ll m_1/2\,,\\[8pt]
		-\frac{2\sqrt{3}}{7}k/W_{-1}\left[-2.2\left(\sqrt{\frac{\gs}{10}}\frac{M_5^3}{\mathcal{C} \, r_c\,M_P \, k^3}\right)^{2/7}\right] &\qquad\text{ for }\Trh\simeq m_1/2\,,\\[8pt]
		6.7\left(\frac{\gs}{10}\right)^{1/4}\left(\frac{M_5^3}{M_P\, r_c\,\mathcal{C}\, k}\right)^{1/2} \,  &\qquad\text{ for }\Trh\gg m_1/2\,,
	\end{cases}
\end{equation}
where $W_{-1} [x]$ corresponds to the $-1$ branch of the Lambert function computed at $x$.

\begin{figure}[t]
	\begin{center}
		\includegraphics[height=0.5\textwidth]{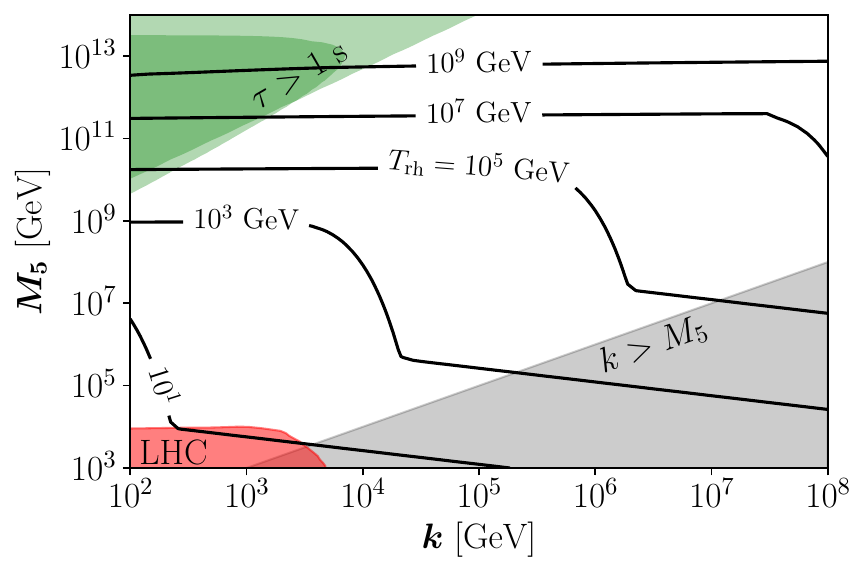}
		\caption{Direct freeze-in: contours for the reheating temperature $\Trh$ required to reproduce the observed DM abundance, for $m_\text{DM}=1$~MeV.
		The red-shaded region is in tension with LHC results from Refs.~\cite{CMS:2018thv,Giudice:2017fmj}.
		The gray-shaded region for which $k > M_5$ is beyond the EFT approach.
		The green-shaded region corresponds to lifetimes of the radion and  the lightest KK-mode longer than 1~s.
}
		\label{fig:direct}
	\end{center}
\end{figure}

Figure~\ref{fig:direct} shows contours for the reheating temperature required to reproduce the observed DM abundance, for $m_\text{DM}=1$~MeV.
In order to compute $\Trh$, the DM yield has been held fixed so that $m_\text{DM}\,Y_0 = \Omega_\text{DM} h^2 \, \frac{1}{\mathfrak{s}_0}\,\frac{\rho_c}{h^2} \simeq 4.3 \times 10^{-10}$~GeV, where $\rho_c \simeq 1.1 \times 10^{-5} \, h^2$~GeV/cm$^3$ is the critical energy density, $\mathfrak{s}_0\simeq 2.9\times 10^3$~cm$^{-3}$ is the entropy density at present and $\Omega_\text{DM} h^2\simeq 0.12$~\cite{Aghanim:2018eyx}.
The gray-shaded region on the lower-right corner corresponding to $k>M_5$ is beyond our EFT approach.
The red-shaded region in the lower-left corner represents the present non-resonant LHC Run-II bounds. The results from non-resonant searches, in contrast with the resonant searches
(see, for example, Refs.~\cite{Aaboud:2017buh, Aaboud:2017yyg}), are not easy to translate into a bound in the $(M_5,\,k)$ plane. In the present work we use the results obtained in Ref.~\cite{CMS:2018thv}, taking advantage of the dedicated analysis performed in Ref.~\cite{Giudice:2017fmj}.
Eventually, the upper-left green corner corresponds to a region of the parameter space for which the lifetime of the radion and of the first KK-graviton are larger than 1 s, 
potentially problematic for BBN (all others KK-modes are heavier and, therefore, will naturally have shorter lifetimes).


\subsection{A rough comparison of CW/LD and RS results}
\begin{figure}[t]
	\begin{center}
		\includegraphics[height=0.5\textwidth]{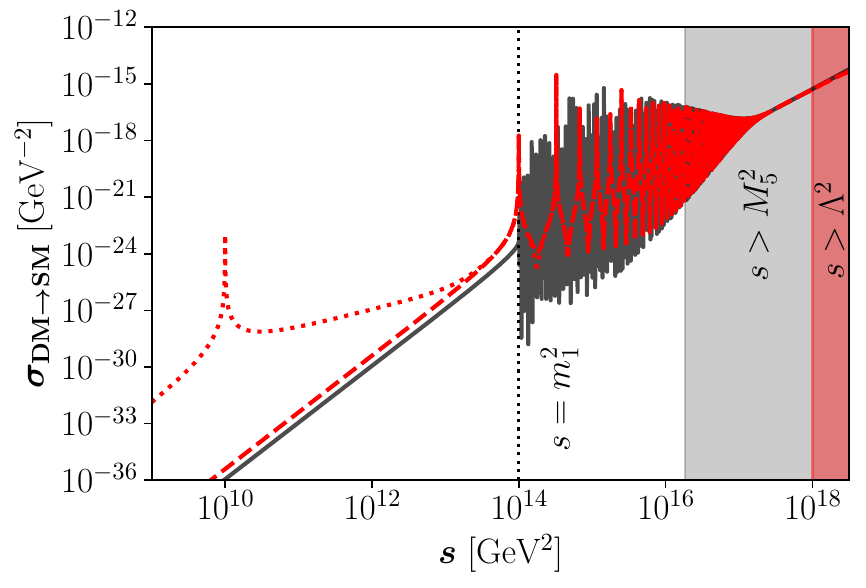}
		\caption{
		The black solid line represents the DM annihilation cross-section in the CW/LD scenario for $k=10^7$~GeV, $\Lambda_{\rm eff}=10^9$~GeV (that corresponds to $M_5 = 1.35721 \times 10^8$ GeV)
		and $m_\text{DM}=1$~MeV.
		The red-dashed (red-dotted) line represents the same process in the RS scenario, for $m_1 = 10^7$ GeV, $\Lambda = 10^9$, $m_\text{DM}=1$~MeV and a radion mass $m_r = 10^7$ GeV ($m_r = 10^5$ GeV).
		The gray- and red-shaded regions on the right of the picture represent the region of the parameter space that is beyond the EFT approach, as $s$ is larger than the scale of the effective theory
		($M_5$ and $\Lambda$ in the case of the CW/LD or the RS scenarios, respectively).
		} \label{fig:crosscheck}
	\end{center}
\end{figure}

We will see throughout all of this Section
that results in the CW/LD model are qualitatively very similar to those obtained in the
RS scenario presented in Ref.~\cite{Bernal:2020fvw}. Even though the CW/LD and the RS scenarios differ fundamentally  in the presence of the KK dilaton tower in the former,  we find that the contribution of the virtual exchange of the dilaton modes to the  DM production via scattering of SM particles is subleading (as it is also the case when the radion mass in RS is similar to the first KK graviton), so one can somehow expect comparable outcomes. We can actually go a step further in our understanding of differences and similarities between the two extra-dimensional scenarios.
In fact, the results of the CW/LD scenario can be quantitatively reproduced by computing the cumulative effect of summing over many KK-states. In order to do so, we find it useful to write eq.~(\ref{virtual_exchange}) 
in terms of an effective coupling $\Lambda_{\rm eff}$ and the mass of the first KK-graviton, $m_1$. First, remember that:
\begin{equation}
	\left \{
	\begin{array}{lll}
	\frac{1}{\Lambda_n^2} & = & \frac{1}{\Lambda_1^2} \times \left [ \frac{n^2 \, (1 + k^2 \, r_c^2)}{n^2 + k^2 \, r_c^2} \right ] \, , \\
	&& \\
	m_n^2 & = & m_1^2 \times \left [ \frac{n^2 + k^2 \, r_c^2}{1 + k^2 \, r_c^2} \right ] \, .
	\end{array}
	\right .
\end{equation}
Using these variables, we find in the regime $T \ll m_1$ (and neglecting the radion/KK-dilaton contribution, that is subleading in the CW/LD scenario as explained above):
\begin{equation}
	\sigma_\text{DM$\to$SM}(s)\Big|_{T \ll m_1} \simeq \frac{49 \, s^3}{1440 \pi} \frac{1}{\Lambda_1^4 m_1^4}  
	\left| \sum_{n=1}^\infty\frac{n^2}{\left ( n^2 + k^2 r_c^2 \right )^2}\right|^2 
	= \frac{49 \, s^3}{1440 \pi} \frac{1}{\Lambda_1^4 m_1^4} \mathcal{F} (k \, r_c) \, ,
\end{equation}
where
\begin{equation}
 \mathcal{F} (k \, r_c)  \equiv \frac{\pi^2}{64} \frac{\left (1 + k^2 \, r_c^2 \right )^4}{k^2 \, r_c^2} \,\mathcal{C} (k \, r_c)\, ,
\end{equation}
being ${\cal C} (k \, r_c)$ the function defined in eq.~(\ref{eq:Cfunction}).
If we now introduce the following effective scale:
\begin{equation}
\label{eq:lambdaeff}
\frac{1}{\Lambda_{\rm eff}} = \frac{1}{\Lambda_1} \mathcal{F} (k \, r_c)^{1/4} \, ,
\end{equation}
it is easy to compare our results in the CW/LD space-time with those obtained in the RS setup in Ref.~\cite{Bernal:2020fvw}. 
This is done in Fig.~\ref{fig:crosscheck} where we compare, as an example,
the annihilation cross-section ${\rm DM} \to {\rm SM}$ from eq.~(\ref{virtual_exchange}) in the RS (red-dotted and red-dashed lines) and CW/LD (black solid line) scenarios for the direct freeze-in. 
The parameters of the two models have been chosen  so as to represent the same physical point in the parameter space: in the case of CW/LD, we choose $k=10^7$ GeV, $\Lambda_{\rm eff}=10^9$~GeV 
(corresponding to $M_5 = 1.35721 \times 10^8$ GeV) and $m_\text{DM}=1$~MeV; for RS, we have $m_1 = 10^7$ GeV, $\Lambda = 10^9$~GeV and $m_\text{DM}=1$~MeV.
In order to compare the two models properly, we have considered two cases for the RS scenario: in the first one (red-dashed line) the radion mass has been chosen identical to the first KK-graviton one, $m_r = m_1 = 10^7$ GeV.
On the other hand, in the second case (red-dotted line) the radion mass is significantly lower than the first KK-graviton mass, $m_r = 10^5$ GeV. 
Remind that in the RS case the radion may have a significant phenomenological impact, in principle, as its mass is an additional free parameter of the model whereas, on the other hand, 
in the CW/LD case the radion mass is not a free parameter and it is tightly linked to the KK-graviton mass. 

The impact of the radion in the RS cross-section can be seen by comparing the two RS cases: for a heavy radion, its contribution is subleading (as it is always the case for CW/LD).
We can see clearly how the annihilation cross-section in the two models coincide both at the qualitative and quantitative level, with a rather small mismatch for low $s$ due to the difference between the first
KK-graviton mass in the two models (that is, $m_1 = \sqrt{k^2 + 1/r_c^2}$ in the CW/LD model and $m_1 = k \, x_1 \Lambda/M_P = 10^7$ GeV in the RS model). Notice how the two resonating patterns
start at $s = m_1^2$ in both cases and, for large $s$, overlap completely.
On the other hand, for a light radion the RS model develop an additional peak for $s \sim m_r^2$ and, in the region in between the radion peak and the starting of the KK-gravitons resonant pattern at $s \sim m_1^2$, 
we can see a much softer evolution of the cross-section with $s$ (that can also be seen in Fig. 1 (left panel) of Ref.~\cite{Bernal:2020fvw}).

The definition of $\Lambda_{\rm eff}$ in eq.~(\ref{eq:lambdaeff}) establishes a relation between the effective scale in RS scenarios, $\Lambda$, and the fundamental parameters of the CW/LD one, $k$ and
$M_5$. We get (apart from numerical factors depending on the combination $k r_c$, that may account for up to one order of magnitude approximately) 
that $\Lambda_{(\rm RS)} \propto \sqrt{M_5^3/k}$ is a rather good approximation to get an understanding of the qualitative features of the CW/LD results as compared to RS ones.
We have indeed found that similar results in the two scenarios can be obtained replacing in the RS formul{\ae} (both for direct and sequential freeze-in) 
$\Lambda_{(\rm RS)} \to \sqrt{M_5/k}$ and $m_1^{(\rm RS)} \to k$. These approximated relations suffice to explain the striking similarities found in the results in the two scenarios.

Notice that our understanding of the relation between RS and CW/LD frameworks allow us to compare easily also Fig.~\ref{fig:cross} (right) with the right panel of Fig.~1 of Ref.~\cite{Bernal:2020fvw}, where the 
same quantity was computed in the framework of a Randall-Sundrum warped extra-dimensional model. Whereas in the CW/LD model we were able to distinguish three regions (approximately below, at and above $m_1$), 
in the case of the RS scenario five distinct regions could be detected. This is a characteristic consequence of the fact that in the RS model the radion mass is an independent
parameter of the model, contrary to the CW/LD case, in which radion and KK-graviton masses are strongly related (see eqs.~\eqref{eq:KKdilatonmasses} and~\eqref{eq:KKgravmasses}, respectively). 
For this reason, in the RS scenario there may be a region in which we have a ``resonant'' behavior with the radion mass and a separated one with the first KK-graviton mass, if the two are far apart one from each other. 
In the CW/LD case, this is not possible and, thus, there is only one intermediate region.

\subsection{Sequential Freeze-in}
\label{sec:sequentialFI}

The second mechanism studied in the present work is the sequential freeze-in, where DM is produced by the decay of bulk particles that have been previously produced via freeze-in.
There are two possibilities to generate the bulk particles: via 2-to-2 annihilations  or via inverse decays of SM states. The two possibilities will be reviewed in detail in Secs.~\ref{sec:sequanni} and~\ref{sec:sequdecay}, respectively.

\subsubsection{Via Annihilations}
\label{sec:sequanni}

The first mechanism that we study in this section is the DM production by the decay of KK-gravitons and radion/KK-dilatons previously produced by SM annihilation processes. 
The Boltzmann eqs.~\eqref{eq:cosmo1}-\eqref{eq:cosmo2} can be simplified in this case as:
\begin{eqnarray}\label{eq:BEFIsecann}
	\frac{dY}{dT}&\simeq&\frac{\gamma_\text{$G_n\to$SM}}{H\,\mathfrak{s}\,T}\left[\left(\frac{Y_G}{Y_G^\text{eq}}\right)^2-1\right]\,\text{BR}(\text{$G_n \to$DM})
	              + \frac{\gamma_{v_n \to\text{SM}}}{H\,\mathfrak{s}\,T}\left[\left(\frac{Y_v}{Y_v^\text{eq}}\right)^2-1\right] \, \text{BR}(v_n \to \text{DM}) \nonumber\\
	&\simeq&-\frac{1}{H\,\mathfrak{s}\,T}\left[\gamma_\text{$G_n\to$ SM} \, \text{BR}(\text{$G_n\to$DM})+\gamma_\text{$v_n  \to$SM} \, \text{BR}(\text{$v_n  \to$DM}) \right] \, ,
\end{eqnarray}
and the interaction rates can be computed using the annihilation cross-section given in Ref.~\cite{Bernal:2020fvw} with the appropriate suppression scale.
Their final form are given by:
\begin{eqnarray}
\label{eq:GammaK-SM}
	\gamma_\text{$G_n\to$SM}(T)&\simeq& 4.8\times 10^4\frac{T^{16}}{\Lambda_{n}^4\,m_n^8} \,, \\
	\gamma_\text{$v_n\to$SM}(T)&\simeq& 7.9\times 10^{-3}\frac{T^{8}}{\Lambda_{\Phi_n}^4} \, , 
	\label{eq:Gammaannr-SM}
\end{eqnarray}
where the index $n$ refers to the specific KK-mode of the two towers of KK-gravitons and radion/KK-dilatons with couplings $1/\Lambda_n$ and $1/\Lambda_{\Phi_n}$, respectively.
The contribution of the KK-dilatons are always negligible with respect to the contribution of the KK-gravitons.  This is a consequence of KK-gravitons being spin-2 massive particles, with 5 polarizations. 
The sum over the polarization states of the KK-graviton gives terms with graviton masses in the denominator of the interaction rate that are compensated by powers of the temperature $T$
in the numerator, thus giving a characteristic $T^{16}$ dependence.  
On the other hand, radion/KK-dilatons are scalar particles and have no polarization. No powers of the radion/KK-dilaton mass appears in the denominator, and the $T$-dependence
of the numerator is proportional to $T^8$, only.  This can be seen comparing eq.~\eqref{eq:Gammaannr-SM} with eq.~\eqref{eq:GammaK-SM}.

The branching ratios that appear in eq.~\eqref{eq:BEFIsecann} can be expressed as:
\begin{equation}
\label{eq:BRKK}
\left \{
\begin{array}{lll}
	\text{BR} (\text{$G_n\to$DM}) & \simeq & \frac{z_n}{z_n+256}\,, \\
	&&\\
	\text{BR} (\text{$v_n \to$DM})    & \simeq & \frac{z_{\Phi_n}}{z_{\Phi_n}+37}\,, 
\end{array}
\right .
\end{equation}
where 
\begin{equation}
\left \{
\begin{array}{lll}
z_n & \equiv & \left (1-4\frac{m_{\text{DM}}^2}{m_n^2}\right )^{5/2} ,\\
&&\\
z_{\Phi_n} & \equiv & \sqrt{1 - 4 \frac{m_{\text{DM}}^2}{m_{v_n}^2}} \left ( 1 + 2 \frac{m_{\text{DM}}^2}{m_{v_n}^2} \right )^{2} .
\end{array}
\right .
\end{equation}
In eq.~\eqref{eq:BRKK} we assume that all SM masses are negligible with respect to the process energy. The only mass that we considered in the analysis is the DM particle mass. 
A description of the bulk particles decays can be found in App.~\ref{App:decays}.

We can use eq.~\eqref{eq:Ydirect}, such as in the direct freeze-in case, in order to obtain the current DM abundance. 
\begin{eqnarray}\label{eq:yield_sfi_ann}
	Y_0 &\simeq& \frac{9.5\times 10^3}{\gss}\sqrt{\frac{10}{\gs}}M_P\,\Trh^{11}\sum_{m_n<\Trh}\frac{1}{\Lambda_n^4 m_n^8}\frac{z_n}{z_n+256}\nonumber\\
	&&+ \frac{5.8\times 10^{-3}}{\gss}\sqrt{\frac{10}{\gs}}M_P\,\Trh^{3}\sum_{m_{v_n}<\Trh}\frac{1}{\Lambda_{\Phi_n}^4}\frac{z_{\Phi_n}}{z_{\Phi_n}+37}\, .
\end{eqnarray}
This expression takes into account all of the KK-gravitons and KK-dilatons with masses below the reheating temperature. We found that these sums can be approximated by
\begin{eqnarray}\label{eq:yield_sfi_ann_parte_2}
	Y_0 &\simeq& \frac{6.9\times 10^{-2}}{\gss}\sqrt{\frac{10}{\gs}}\frac{M_P\,\Trh^{11}}{M_5^6\,k^7\,r_c}\mathcal{H}(\Trh-m_1)\nonumber\\
	&&+\frac{1.2\times 10^{-7}}{\gss}\sqrt{\frac{10}{\gs}}\frac{M_P\,\Trh^{3}\,k^2}{M_5^6}\mathcal{H}(\Trh-m_{v_1})\,,
	\end{eqnarray}
where $\mathcal{H}$ is the Heaviside step function.
As for the direct freeze-in case, the bulk particles must be out of chemical equilibrium with the thermal bath. This condition is equivalent to impose that $\gamma_\text{$G_n\to$SM} \ll n_\text{eq}\,H$ 
and $\gamma_\text{$\Phi_n  \to$SM} \ll n_\text{eq}\,H$. Eventually, avoiding chemical equilibrium translates into a bound over the reheating temperature:
\begin{equation}
	\Trh \ll  0.3\left[\sqrt{\frac{\gs}{10}}\frac{\Lambda_1^4\,m_1^8}{M_P}\right]^{1/11} \, .
\end{equation}

\begin{figure}[t]
	\begin{center}
		\includegraphics[height=0.32\textwidth]{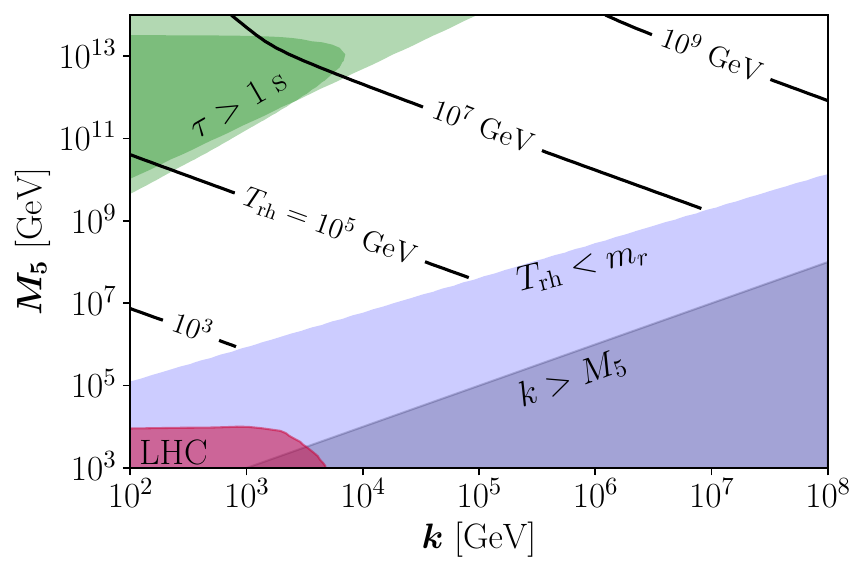}
		\includegraphics[height=0.32\textwidth]{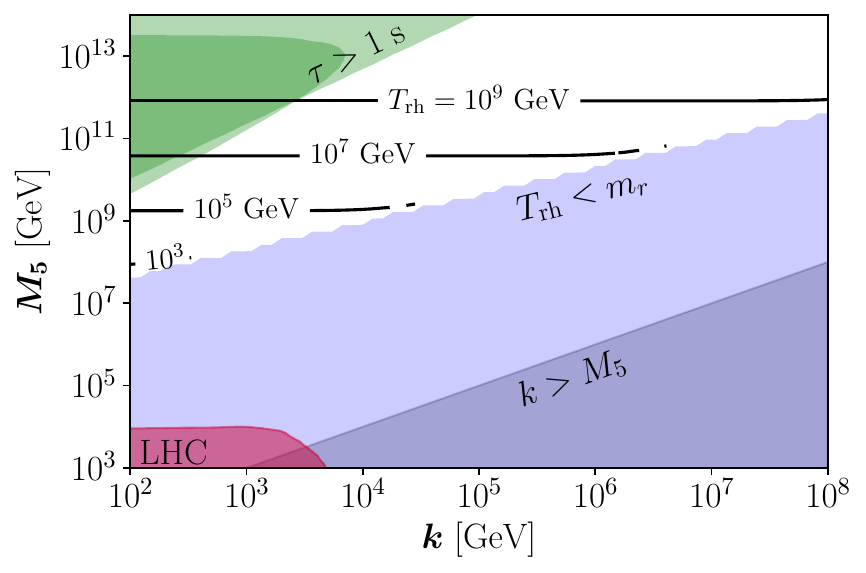}
		\caption{Sequential freeze-in via annihilations (left panel) and via inverse decays (right panel): contours for the reheating temperature 
		required to reproduce the observed DM abundance, for $m_\text{DM}=1$~MeV.
	         The red-shaded region is in tension with LHC Run-II results from Refs.~\cite{CMS:2018thv,Giudice:2017fmj}.
		The dark blue-shaded region for which $k > M_5$ is beyond the EFT approach.
		The green-shaded region corresponds to lifetimes of KK-modes longer than 1~s.
		Eventually, the light blue-shaded region depicts the region of the parameter space for which $\Trh < m_{v_1}$.
}
		\label{fig:seq_FI_ann}
	\end{center}
\end{figure}

In the left panel of Fig.~\ref{fig:seq_FI_ann} we summarized the results for the sequential freeze-in via 2-to-2 processes. We show in this figure the value of the 5-dimensional Planck mass $M_5$  needed to obtain the current value of the DM relic abundance, as a function of the first KK-graviton mass, $m_1$. We notice that sequential freeze-in is only viable if $\Trh > m_{v_1}$, as KK-states have to be produced in the thermal plasma.
As in Fig.~\ref{fig:direct}, the red-shaded region represent the present LHC bounds, while the green-shaded area 
is the region of the parameter space where the radion lifetime is too long and produce problems with BBN. The gray-shaded region represents the $k >  M_5$ area, 
where the EFT description cannot be applied (as in this region the first KK-graviton mass, $m_1$, is larger than the fundamental scale of the theory, $M_5$). Eventually, the light blue-shaded area
is the region of the parameter space for which the reheating temperature $\Trh$ is smaller than the radion mass.

\subsubsection{Via Inverse Decays}
\label{sec:sequdecay}

The last mechanism that we study in the present work is the sequential freeze-in via inverse decay. In this mechanism the DM abundance is consequence of bulk particles annihilation, previously produced via inverse decay of SM particles. The Boltzmann equation in this case reduces to
\begin{eqnarray}\label{eq:BEFIsecdec}
	\frac{dY}{dT}&\simeq&\frac{\gamma^d_\text{$G_n\to$SM}}{H\,\mathfrak{s}\,T}\left[\frac{Y_G}{Y_G^\text{eq}}-1\right]\,\text{BR}(\text{$G_n\to$DM})
	              + \frac{\gamma^d_\text{$v_n \to$SM}}{H\,\mathfrak{s}\,T}\left[\frac{Y_v}{Y_v^\text{eq}}-1\right]\,\text{BR}(\text{$v_n \to$DM}) \nonumber\\
	&\simeq&-\frac{1}{H\,\mathfrak{s}\,T}\left[\gamma^d_\text{$G_n\to$SM}\,\text{BR}(\text{$G_n\to$DM}) + \gamma^d_\text{$v_n \to$SM}\,\text{BR}(\text{$v_n \to$DM}) \right] \, .
\end{eqnarray}
Using the expression defined in Ref.~\cite{Bernal:2020fvw}  and taking the decay from App.~\ref{App:decays}, it is possible to obtain the interaction rate densities for KK-gravitons and KK-dilatons:
\begin{eqnarray}
        \gamma^d_\text{$G_n\to$SM}&\simeq&\frac{73}{480\pi^3} \frac{m_n^5\,T}{\Lambda_n^2}\,K_1\left(\frac{m_n}{T}\right),\\
	\gamma^d_\text{$v_n \to$SM}&\simeq&\frac{111}{32\pi^3} \frac{m_{v_n}^5\,T}{\Lambda_{\Phi_n}^2}\,K_1\left(\frac{m_{v_n}}{T}\right).
\end{eqnarray}
The equations above present a significant difference with eqs.~\eqref{eq:GammaK-SM} and~\eqref{eq:Gammaannr-SM}: in this case, the interaction rate comes from the decay of the bulk particles into SM particles
and, as a consequence, there is no enhancement due to graviton polarization.
As in the other two cases, eq.~\eqref{eq:Ydirect} allows to obtain the current DM relic abundance
\begin{equation}\label{eq:YieldSecInv1}
	Y_0\simeq\frac{5.6\times 10^{-2}}{\gss}\sqrt{\frac{10}{\gs}}M_P\sum_{m_n<\Trh}\frac{m_n}{\Lambda_n^2}\frac{z_n}{z_n+256} + \frac{1.3}{\gss}\sqrt{\frac{10}{\gs}}M_P\sum_{m_{v_n}<\Trh}\frac{m_{v_n}}{\Lambda_{\Phi_n}^2}\frac{z_{\Phi_n}}{z_n+37}\, .
\end{equation}
In this case, we can see numerically that the most part of the production takes place at
$T \simeq m/ 2.5$, where $m$ corresponds to the mass of the lightest KK-dilatons and KK-gravitons.
This fact indirectly introduces a dependence on $\Trh$, even if the terms in the sums of the above expression do not present any explicit dependence on it. With this information, and taking into account bulk states with masses below the reheating temperature, only, 
we can write the above expression as:
\begin{eqnarray}\label{eq:YieldSecInv2}
	Y_0&\simeq&\frac{3.5\times 10^{-5}}{\gss}\sqrt{\frac{10}{\gs}}\frac{M_P}{M_5^3}\left(\Trh^2-m_1^2\right)\mathcal{H}(\Trh-m_1) \nonumber\\
	&&\quad + \frac{5.1\times 10^{-3}}{\gss}\sqrt{\frac{10}{\gs}}\frac{M_P\,k^2}{M_5^3}\left[\frac{\sqrt{2}}{8}\mathcal{H}(\Trh-m_{v_1})+\frac{1}{\pi}\log\frac{\Trh}{m_{v_1}}\mathcal{H}(\Trh-m_{v_1})\right].\qquad
\end{eqnarray}

Eventually, as for other cases discussed above, we must be sure that KK-gravitons and radions/KK-dilatons never reach the equilibrium with the thermal bath (that is, 
$\gamma^d_\text{$G_n\to$SM} \ll n_\text{eq}\,H$ and $\gamma^d_\text{$\Phi_n  \to$SM} \ll n_\text{eq}\,H$). We may, therefore, derive a bound on  the reheating temperature:
\begin{equation}
	\Trh \ll  0.34 \left [ \sqrt{\frac{10}{\gs}} \, \frac{M_P\,m_1^4}{\Lambda_1^2}\right]^{1/3}   \,     .
\end{equation}
In the right panel of Fig.~\ref{fig:seq_FI_ann} we can see the different results of the sequential freeze-in due to inverse decays. 
The red-, green- and gray-shaded regions  in the $(M_5, m_1)$ plane represent, respectively, the bounds from non-resonant searches at LHC, the BBN bounds due to the dilaton 
decays and the limit of the EFT theory. The light blue-shaded area, eventually, represents the region of the parameter space for which $m_{v_1} \geq \Trh$. 

\subsection{A comprehensive analysis of the results for direct and sequential freeze-in}
\label{sec:summary}

\begin{figure}[t]
	\begin{center}
		\includegraphics[height=0.5\textwidth]{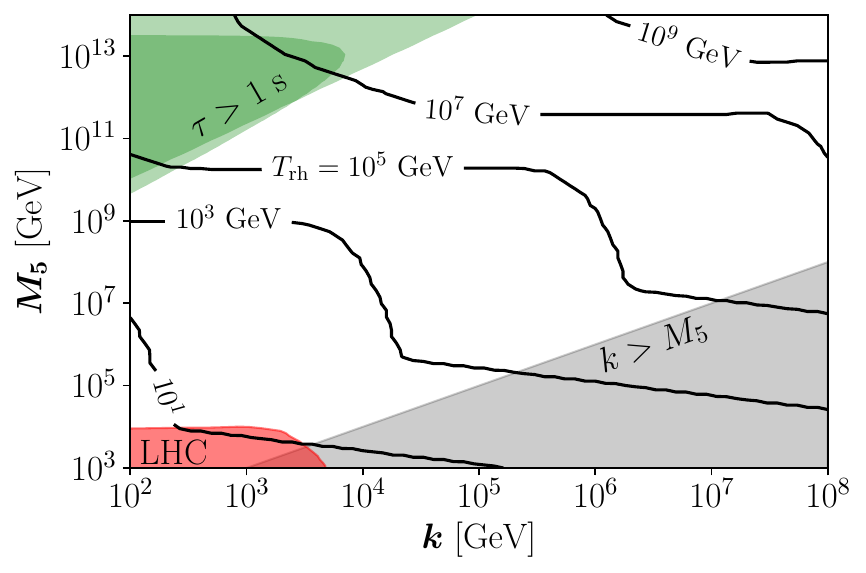}
		\caption{ Contours for the reheating temperature $\Trh$ required to reproduce the observed DM abundance, for $m_\text{DM}=1$~MeV.
		 The red-shaded region is in tension with LHC Run-II results  from Refs.~\cite{CMS:2018thv,Giudice:2017fmj}.
		The gray-shaded region for which $k > M_5$ is beyond the EFT approach.
		The green-shaded region corresponds to lifetimes of the lighter KK-mode longer than 1~s.
}
		\label{fig:final}
	\end{center}
\end{figure}

In Fig.~\ref{fig:final} we eventually put together the results of direct and sequential freeze-in (including both annihilation and inverse decays), from Secs.~\ref{sec:directFI} and~\ref{sec:sequentialFI}.
The plot represents the reheating temperature $\Trh$ for a scalar DM particle with mass $m_{\rm DM} = 1$ MeV, as a function of the other two free parameters of the CW/LD scenario, $k$ and $M_5$.
The red-shaded region in the lower-left corner represents the portion of the parameter space excluded by present LHC Run-II results from Refs.~\cite{CMS:2018thv,Giudice:2017fmj}. 
The gray-shaded area in the lower-right corner is the region where we cannot trust our effective theory approach, as the mass of the KK-gravitons ($\propto k$) is larger than the fundamental scale of the model, $M_5$. Eventually, the green-shaded area in the upper-left corner represents the region of the parameter space for which the lifetime of the lighter KK-mode is larger than 1~s (a case for which we could have problems 
for a successful BBN). 
Differently from the case of the freeze-out mechanism in extra-dimensions (studied in Refs.~\cite{Folgado:2019sgz,Folgado:2019gie,Folgado:2020vjb} both in the RS and CW/LD scenarios), 
where the LHC (in its present form or in its high-luminosity upgrade) may constrain significantly the allowed parameter space, we can see that for freeze-in in the CW/LD model the region of the parameter space 
tested by the LHC Run-II is quite limited. This was also true in the case of the RS scenario (see Fig.~6 of Ref.~\cite{Bernal:2020fvw}). In both cases, we have found that a very large region of the parameter space 
is not excluded by neither theoretical nor experimental constraints. In the case of the CW/LD scenario, discussed here, we see that the reheating temperature needed for a successful freeze-in can range from 
$\Trh$ as low as 10 GeV to $\Trh \sim 10^9$ GeV for $k \in [10^2, 10^8]$ GeV and $M_5 \in [10^4,10^{13}]$ GeV (in the case of RS, $\Trh$ could be as low as 1 GeV for $m_1 \sim 1$ TeV and $\Lambda \sim 10^8$ GeV). 
Particularly interesting is the lower-left corner of the picture where, once LHC Run-II bounds are
taken into account, we see that the observed DM relic abundance can be obtained for $M_5 \sim 10$ TeV, $k \in [0.1, 10]$ TeV and $\Trh$ ranging from 10 to 100 GeV, approximately. In this region we are able 
to achieve a rather small hierarchy between the fundamental scale of gravity, $M_5$, and the electro-weak scale, $\Lambda_{\rm EW}$.

%% file: appendix.tex
\section{Bulk particle Decays}
\label{App:decays}

KK-gravitons can decay into both SM and DM particles. The corresponding decay widths are:
\begin{eqnarray}
	\Gamma_{G_n\to\text{SM}} &\simeq& \frac{73}{240\, \pi} \frac{m_n^3}{\Lambda_n^2} \, ,\label{eq:decKKtoSM}\\
	\Gamma_{G_n\to\text{DM}} &=& \frac{m_n^3}{960 \, \pi \Lambda_n^2} \left ( 1 - 4\frac{m_\text{DM}^2}{m_n^2} \right )^{5/2},\label{eq:decKKtoDM}
\end{eqnarray}
where all SM masses were neglected for simplicity.
Eventually, the decay widths of dilatons into SM and DM particles are:
\begin{eqnarray}
	\Gamma_{v_n\to\text{SM}} &\simeq& \frac{37 m_{v_n}^3}{32 \, \pi \Lambda_{\Phi_n}^2} \,,\label{eq:decrtoSM}\\
	\Gamma_{v_n\to\text{DM}} &=& \frac{m_{v_n}^3}{32 \, \pi \Lambda_{\Phi_n}^2} \left ( 1 - 4\frac{m_\text{DM}^2}{m_{v_n}^2} \right )^\frac12 \left (  1 +2\frac{m_\text{DM}^2}{m_{v_n}^2} \right )^{2}   \, ,
	\label{eq:decrtoDM}
\end{eqnarray}
where again all SM masses were neglected for simplicity.

\section{Example of the sums over KK-gravitons}
\label{app:kksum}

In this Appendix we show an example of how the interaction rate densities are analytically computed.
For instance, the contribution to the DM annihilation cross-section into SM states coming from the exchange of KK-gravitons in eq.~\eqref{virtual_exchange} is
\begin{equation}
	\sigma_\text{DM$\to$SM}(s) \simeq \frac{49 \, s^3}{1440 \, \pi} \left|\sum_{n=1}^\infty\frac{1}{\Lambda_n^2}\frac{1}{s-m_n^2+i\,m_n\Gamma_n}\right|^2 \equiv \frac{49 \, s^3}{1440 \pi} \left|S_\text{KK}\right|^2 \, .
\end{equation}
In the limit $s \ll m_1^2$,
\begin{equation}
	S_\text{KK} \simeq \sum_{n=1}^\infty \frac{1}{\Lambda_n^2\,m_n^2} \simeq \frac{1}{8\,M_5^3\,k}\,\frac{\sinh(2\pi k\,r_c)-2\pi k\,r_c}{\sinh^2(\pi k\,r_c)} = \frac{1}{8\,M_5^3\,k}\, {\cal C}^{1/2} (k r_c) \, ,
\end{equation}
where eqs.~\eqref{eq:KKgravmasses} and~\eqref{Lambda_graviton} were used.

One typically needs to compute interaction rate densities $\gamma$ as a function of the temperature $T$.
In general, for the process where two particles ($i$, $j$) annihilate into two states ($k$, $l$), the interaction rate density $i + j \rightarrow  k + l$ is defined as:
\begin{equation}\label{eq:gamma_sin_approx}
        \gamma(T) \equiv \frac{T}{64\,\pi^4} \int_{s_\text{min}}^{s_\text{max}}ds \sqrt{s}\,\sigma_R(s)\,K_1\left(\frac{\sqrt{s}}{T}\right)\,,
\end{equation}
where $s_{\text{min}}\equiv \text{max}\left[(m_i + m_j)^2,\, (m_k + m_l)^2\right]$, $s_\text{max}$ corresponds to the center-of-mass energy until which the theory is valid, $\sigma_R$ is the reduced cross-section summed 
over all the degrees of freedom of the initial and final states, and $K_1$ is a modified Bessel function.
$\sigma_R$ corresponds to the total cross-section $\sigma(s)$ without the flux factor, and can be written as:
\be
\sigma_{R}(s) = 2\,\frac{\left[s-(m_i+m_j)^2\right]\left[s-(m_i-m_j)^2\right]}{s}\,\sigma(s)\,.
\ee
In the present case, the interaction rate density $\gamma_\text{DM$\to$SM}(T)$ is expressed as a function of the reduced cross-section $\sigma_R(s) \simeq 2s\,\sigma_\text{DM$\to$SM}(s)$, and is given by
\begin{eqnarray}
	\gamma_\text{DM$\to$SM}(T) &\simeq& \frac{T}{64\, \pi^4}\int_0^{M_5^2} \sigma_R\,\sqrt{s}\,K_1\left(\frac{\sqrt{s}}{T}\right)\,ds
	\simeq \frac{49 \, T}{46080\, \pi^5}| \, S_\text{KK}|^2 \, \int_0^{M_5^2} s^{9/2} \,K_1\left(\frac{\sqrt{s}}{T}\right)\,ds\nonumber\\
	&\simeq& 1.7\times 10^{-1}\frac{T^{12}}{M_5^6\,k^2}\left[\frac{\sinh(2\pi k\,r_c)-2\pi k\,r_c}{\sinh^2(\pi k\,r_c)}\right]^2,
\end{eqnarray}
for the case $T\ll m_1/2$, as reported in eq.~\eqref{eq:gammaann}.
Remind that the integral stops at $s=M_5^2$, as it is the limit of validity of the EFT considered.
For the case $T\gg m_1/2$, the analytical computation is much more involved as it requires integrating over a huge number of resonances.
All the analytical approximations have been validated by comparing with the numerical calculations.